\documentclass{article}
\usepackage{cite}
\usepackage{graphicx}
\usepackage{dcolumn}

\input{tcilatex}
\begin{document}

\date{}
\title{The truncated Coulomb potential revisited}
\author{Francisco M. Fern\'{a}ndez\thanks{%
fernande@quimica.unlp.edu.ar} \\
INIFTA, DQT, Sucursal 4, C. C. 16, \\
1900 La Plata, Argentina}
\maketitle

\begin{abstract}
We apply the Frobenius method to the Schr\"{o}dinger equation with a
truncated Coulomb potential. By means of the tree-term recurrence relation
for the expansion coefficients we truncate the series and obtain exact
eigenfunctions and eigenvalues. From a judicious arrangement of the exact
eigenvalues we derive useful information about the whole spectrum of the
problem and can obtain other eigenvalues by simple and straightforward
interpolation.
\end{abstract}

\section{Introduction}

\label{sec:intro}

The family of truncated Coulomb potentials $V(r)=-Ze^{2}/\left(
\beta ^{k}+r^{k}\right) ^{1/k}$, $k=1,2,\ldots $, has received
considerable attention\cite{MP78,LM81,P81,M81, LLALM82, DMV85,
SVD85, RM89, DV90, F91, D94}. In particular, the case $k=2$ proved
to be suitable as an approximation to the laser-dressed binding
potential\cite{LM81}, in the UV and X-ray absorption cross section
by hydrogen in the presence of an intense IR laser field\cite{M81}
and in the discussion of the atomic states of hydrogen in the
presence of intense laser fields\cite{LLALM82}. The
Schr\"{o}dinger equation has been solved in many different ways,
for example, Patil\cite{P81} described some analytic properties of
the scattering phase shifts of a variety of potentials, including
the cases $k=1$ and $k=2$. Dutt et al\cite{DMV85} applied $1/N$
expansion to the model with $k=2$. Singh et al\cite{SVD85}
obtained accurate eigenvalues for the cases $k=1$ and $k=2$ by
means of a transformation of the radial equation followed by some
iterative method. Ray and Mahata\cite{RM89} calculated the
energies of several states for the case $k=1$ by means of the
$1/N$ expansion. De Meyer and Vanden Berghe\cite{DV90} calculated
eigenvalues of the model with $k=1$ from a secular equation
derived by means of the Lie algebra SO(2,1) and a scaling
parameter introduced by means of a canonical transformation.
Fern\'{a}ndez\cite{F91} applied the Frobenius method, derived a
three-term recurrence relation for the expansion coefficients and
obtained exact eigenvalues and eigenfunctions of the model with
$k=1$ by suitable truncation of the series. By means of
sypersymmetry Drigo Filho\cite{D94}
obtained the same three-term recurrence relation derived earlier by Fern\'{a}%
ndez for the case $k=1$ and discussed some eigenvalues and eigenfunctions.

The purpose of this paper is a more detailed analysis of the Frobenius
method. In section~\ref{sec:model} we derive a suitable dimensionless
eigenvalue equation for the radial part of the Schr\"{o}dinger equation with
the truncated Coulomb potential. In section~\ref{sec:TTRR} we discuss and
interpret the distribution of the eigenvalues stemming from the truncation
of the Frobenius series. Finally, in section~\ref{sec:conclusions} we
summarize the main results and draw conclusions.

\section{The model}

\label{sec:model}

Throughout this paper we restrict ourselves to the case $k=1$ and focus on
the model given by Hamiltonian operator
\begin{equation}
H=-\frac{\hbar ^{2}}{2m}\nabla ^{2}-\frac{Ze^{2}}{4\pi \epsilon _{0}\left(
r+r_{0}\right) },  \label{eq:H}
\end{equation}
where $m$ is a reduced mass, $e$ is the electron charge, $Z$ the
atomic number, $\epsilon _{0}$ is the vacuum permitivity and
$r_{0}$ a suitable cut-off radius. It is convenient to work with
dimensionless equations\cite{F20}; for example, if, in the present
case, we define the dimensionless coordinate
$\tilde{\mathbf{r}}=\mathbf{r}/r_{0}$ and the Laplacian $\tilde{\nabla}%
^{2}=r_{0}^{2}\nabla ^{2}$ the Hamiltonian operator (\ref{eq:H}) becomes
\begin{equation}
\tilde{H}=\frac{mr_{0}^{2}}{\hbar ^{2}}H=-\frac{1}{2}\tilde{\nabla}^{2}-%
\frac{\beta }{\tilde{r}+1},\;\beta =\frac{mr_{0}Ze^{2}}{4\pi \epsilon
_{0}\hbar ^{2}},  \label{eq:H_dim_1}
\end{equation}
where the eigenvalues $E$ of $H$ and $\tilde{E}$ of $\tilde{H}$ are related
by $\tilde{E}=mr_{0}^{2}E/\hbar ^{2}$.

In what follows we will focus on solving the radial Schr\"{o}dinger equation
associated to the dimensionless operator $\tilde{H}$; however, for
comparison purposes it is convenient to show the connection between this
operator and those studied in earlier papers. For example, if we now define $%
\breve{r}=\beta \tilde{r}$ and $\breve{\nabla}^{2}=\beta ^{-2}\tilde{\nabla}%
^{2}$ we obtain an alternative expression for the dimensionles Hamiltonian
operator
\begin{equation}
\breve{H}=\beta ^{-2}\tilde{H}=-\frac{1}{2}\breve{\nabla}^{2}-\frac{1}{%
\breve{r}+\beta },  \label{eq:H_dim_2}
\end{equation}
where the energy eigenvalues $\tilde{E}$ of $\tilde{H}$ and $\breve{E}$ of $%
\breve{H}$ are related by $\tilde{E}=\beta ^{2}\breve{E}$.

\section{Exact solutions}

\label{sec:TTRR}

The radial part of the Schr\"{o}dinger equation with the Hamiltonian (\ref
{eq:H_dim_1}) can be written as
\begin{equation}
\left[ -\frac{1}{2}\frac{d^{2}}{dr^{2}}+\frac{l(l+1)}{2r^{2}}-\frac{\beta }{%
r+1}\right] f(r)=Ef(r),\;f(0)=0,\;\lim\limits_{r\rightarrow \infty }f(r)=0,
\label{eq:radial_equation}
\end{equation}
where $l=0,1,\ldots $ is the angular momentum quantum number and we have
omitted the tilde over the radial variable.

In order to solve the eigenvalue equation
(\ref{eq:radial_equation}) we resort to the well known Frobenius
method and instead of the ansatz used earlier\cite{F91} here we
try
\begin{equation}
f(r)=r^{l+1}(1+r)e^{-\alpha r}\sum_{j=0}^{\infty }c_{j}r^{j},\;\alpha =\sqrt{%
-2E}.  \label{eq:f_ansatz}
\end{equation}
A straightforward calculation shows that the expansion coefficients $c_{j}$
satisfy the recurrence relation
\begin{eqnarray}
c_{j+2} &=&A_{j}c_{j+1}+B_{j}c_{j},\;j=-1,0,1,2,\ldots ,\;c_{-1}=0,\;c_{0}=1,
\nonumber \\
A_{j} &=&\frac{2\alpha \left( j+l+2\right) -j^{2}-j\left( 2l+5\right)
-2\left( 2l+3\right) }{\left( j+2\right) \left( j+2l+3\right) },  \nonumber
\\
B_{j} &=&2\frac{\alpha \left( j+l+2\right) -\beta }{\left( j+2\right) \left(
j+2l+3\right) }.  \label{eq:TTRR}
\end{eqnarray}

We obtain exact eigenvalues and eigenfunctions if the truncation conditions $%
c_{n}\neq 0$, $c_{n+1}=c_{n+2}=0$, $n=0,1,\ldots $, are satisfied by real
positive values of $\alpha $ and $\beta $ because, in such a case, $c_{j}=0$
for all $j>n$ and the series in equation (\ref{eq:f_ansatz}) becomes a
polynomial of degree $n$. These truncation conditions are equivalent to
\begin{equation}
B_{n}=0,\;c_{n+1}=0,\;n=0,1,\ldots .  \label{eq:trunc_cond}
\end{equation}
The first equation in (\ref{eq:trunc_cond}) gives us a simple relationship
between $\alpha $ and $\beta $: $\beta =\alpha \left( n+l+2\right) $ and
from $c_{n+1}=0$ we obtain $\alpha $.

For example, for $n=0$ we obtain $\beta =l+2$ and $\alpha =1$ and the
corresponding eigenfunction does not have nodes because $c_{j}=0$ for all $%
j>0$.

For $n=1$ we have
\begin{eqnarray}
\beta _{l}^{(1,i)} &=&\alpha _{l}^{(1,i)}\left( l+3\right) ,\;i=1,2,
\nonumber \\
\alpha _{l}^{(1,1)} &=&\frac{3\sqrt{l+2}-\sqrt{l+6}}{2\sqrt{l+2}}<\alpha
_{l}^{(1,2)}=\frac{\sqrt{l+6}+3\sqrt{l+2}}{2\sqrt{l+2}}.
\label{eq:alpha,beta,n=1}
\end{eqnarray}
From the coeficients
\begin{equation}
c_{1,l}^{(1,1)}=\frac{\sqrt{l+2}-\sqrt{l+6}}{2\sqrt{l+2}},\;c_{1,l}^{(1,2)}=%
\frac{\sqrt{l+6}+\sqrt{l+2}}{2\sqrt{l+2}},  \label{eq:c_1^1}
\end{equation}
we conclude that the ansatz $f_{l}^{(1,1)}(r)$ has one node and $%
f_{l}^{(1,2)}(r)$ is nodeless. This result is consistent with the fact that $%
E_{l}^{(1,1)}>E_{l}^{(1,2)}$, where $E_{l}^{(n,i)}=-\left[ \alpha
_{l}^{(n,i)}\right] ^{2}/2$.

In general we obtain $\beta _{l}^{(n.i)}=\alpha _{l}^{(n,i)}\left(
n+l+2\right) $, $n=0,1,\ldots $, $i=1,2,\ldots ,n+1$, which we arrange so
that $\alpha _{l}^{(n,i+1)}>\alpha _{l}^{(n,i)}$, and the corresponding
eigenfunctions are given by
\begin{equation}
f_{l}^{(n,i)}(r)=r^{l+1}(1+r)e^{-\alpha
_{l}^{(n,i)}r}\sum_{j=0}^{n}c_{j,l}^{(n,i)}r^{j},  \label{eq:f^(n,i)}
\end{equation}
where $f_{l}^{(n,i)}(r)$ has $n+1-i$ nodes. It can be proved that
all the roots of $c_{n+1}=0$ are real, as was already done for
other models\cite{LM86,CDW00, AF20}.

The eigenvalues of the radial equation (\ref{eq:radial_equation}) are
commonly labelled as $E_{\nu ,l}(\beta )$, where $\nu =0,1,\ldots $ in such
a way that $E_{\nu ,l}<E_{\nu +1,l}$ from which we obtain the corresponding
parameters $\alpha _{\nu ,l}(\beta )=\sqrt{-2E_{\nu ,l}}$. The question
arises as to the relation between $\alpha _{l}^{(n,i)}$ and $\alpha _{\nu
,l} $. It follows from the arguments given above that $\left( \beta
_{l}^{(n,i)},\alpha _{l}^{(n,i)}\right) $ is a point on the curve $\alpha
_{\nu ,l}(\beta )$ for $\nu =n+1-i$.

\begin{figure}[tbp]
\begin{center}
\includegraphics[width=9cm]{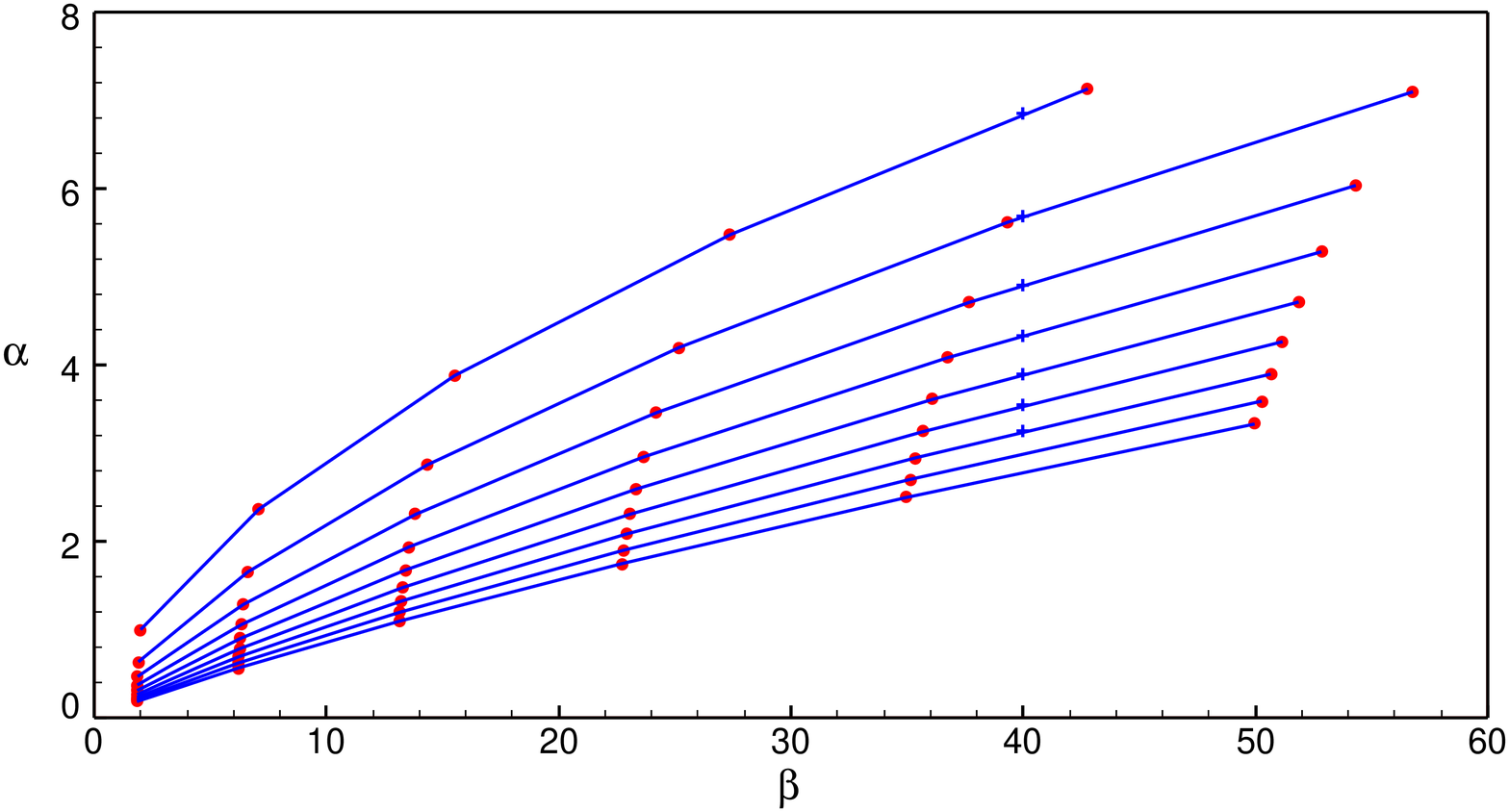}
\end{center}
\caption{Eigenvalues $\alpha_{\nu ,0}$, $\nu=0,1,\ldots,8$, obtained from
the truncation condition (red circles) and numerically (blue crosses)}
\label{fig:alpha_l=0}
\end{figure}

Figure~\ref{fig:alpha_l=0} shows some values of $\alpha $
calculated from the truncation condition with $l=0$ (red circles)
and suitable continuous lines that connect points on the curves
$\alpha _{\nu ,0}$, $\nu =0,1,\ldots ,8$. Numerical values of
$\alpha _{\nu ,0}$, obtained by means of the Riccati-Pad\'{e}
method (RPM)\cite{FMT89a} for $\beta =40$, marked by crosses,
appear on the blue lines between red circles confirming the
argument put forward above. Notice that $E_{\nu ,l}$ decreases
with $\beta $ in agreement with the Hellmann-Feynman theorem
\begin{equation}
\frac{dE}{d\beta }=-\left\langle \frac{1}{r+1}\right\rangle ,  \label{eq:HFT}
\end{equation}
and, consequently, $\alpha _{\nu ,l}$ increases with this parameter.

\begin{figure}[tbp]
\begin{center}
\includegraphics[width=9cm]{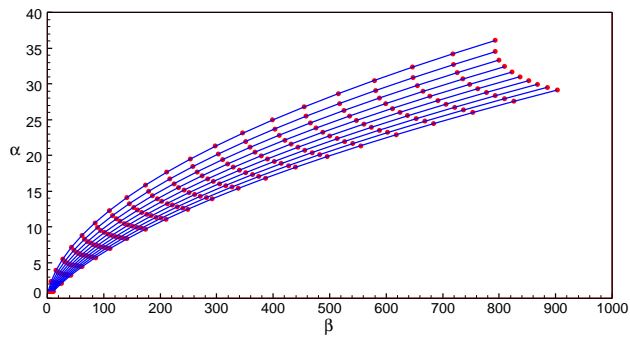}
\end{center}
\caption{Eigenvalues $\alpha_{0 ,l}$, $l=0,1,\ldots,9$, obtained from the
truncation condition}
\label{fig:alpha_l}
\end{figure}

Figure~\ref{fig:alpha_l} shows values of $\alpha _{0,l}$ for $l=0,1,\ldots
,9 $ connected by blue lines along every curve $\alpha _{0,l}(\beta )$. For
every value of $\beta $ $\alpha _{0,l}$ decreases with $l$.

\begin{figure}[tbp]
\begin{center}
\includegraphics[width=9cm]{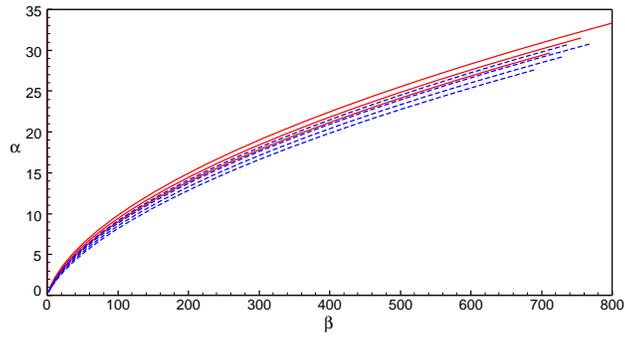}
\end{center}
\caption{Eigenvalues $\alpha_{0 ,2} > \alpha_{1,1} > \alpha_{2,0}$ (red,
continuous line) and $\alpha_{0 ,3} > \alpha_{1,2} > \alpha_{2,1} >
\alpha_{3,0}$ (blue, dashed line) obtained from the truncation condition}
\label{fig:alpha_nl}
\end{figure}

In the case of the Coulomb problem ($r_{0}=0$, or $\beta =0$) all $\alpha
_{\nu ,k-\nu }$, $\nu =0,1,\ldots ,k$ are identical because the energy
depends on $\nu +l$. The curves $\alpha _{\nu ,2-\nu }(\beta )$, $\nu =0,1,2
$ and $\alpha _{\nu ,3-\nu }(\beta )$, $\nu =0,1,2,3$ in Figure~\ref
{fig:alpha_nl} shows how the cut off removes this degeneracy.

The continuous lines in figures \ref{fig:alpha_l=0}, \ref{fig:alpha_l} and
\ref{fig:alpha_nl} are just straight lines that connect pairs of points
given by the truncation condition (\ref{eq:trunc_cond}). In principle, one
can obtain reasonably accurate values of $\alpha $ between those points by
suitable interpolation; for example Lagrange interpolation. Using 21 exact
points of the curve $\alpha _{0,0}(\beta )$ obtained from the truncation
condition we estimate $\alpha _{0,0}^{TC}(40)=6.856$ while the accurate
numerical RPM result is $\alpha _{0,0}^{RPM}=6.854786377$. This accuracy is
obviously sufficient for any physical application of such a simple model.

\section{Conclusions}

\label{sec:conclusions}

The truncated Coulomb model (\ref{eq:H}) is quasi-sovable (or conditionally
solvable) because we can only obtain eigenvalues and eigenfunctions exactly
for some values of the model parameter $\beta $. However, if we arrange the
roots $\beta _{l}^{(n,i)}$, $\alpha _{l}^{(n,i)}$, provided by the
truncation condition, judiciously we derive useful information about the
spectrum of the problem. For example, we can obtain sufficiently accurate
curves $\alpha _{\nu ,l}(\beta )$ or $E_{\nu ,l}(\beta )$ by simple and
straighforward interpolation of the points given by the truncation condition
(\ref{eq:trunc_cond}).

\end{document}